# Dynamic Wettability Modulation of Textured, Soft and LIS Interfaces Using Electrowetting

*Deepak J[a], Suman Chakraborty[b], Shubham S. Ganar[a], and Arindam Das[a]*

[a]School of Mechanical Sciences, Indian Institute of Technology (IIT) Goa, GEC Campus, Farmagudi, Ponda, Goa 403401, India

[b] Department of Mechanical Engineering, Indian Institute of Technology, Kharagpur-721302, India

**Abstract**

Electrowetting on textured and lubricant-infused surfaces is conventionally expected to promote enhanced droplet spreading by reducing apparent contact angles. Contrary to this intuition, we report rapid tangential droplet ejection at applied DC voltages on specific microtextured, lubricant-infused surfaces. Using high-speed imaging and a precisely controlled electrowetting setup, we reveal the dependence of droplet dynamics on surface topology, wetting state, and the presence of a lubricant. On densely textured thick PDMS substrates (post spacing 5–10 μm) in a low-hysteresis non-wetting (Cassie) state, and on all lubricant-infused textured surfaces, droplets experience sudden lateral motion and eventual detachment. We attribute this counterintuitive phenomenon to unbalanced electrocapillary forces at the contact line combined with minimal pinning, which allows asymmetries in electric stresses to translate directly into net lateral motion. In contrast, Wenzel-state droplets or surfaces with larger texture spacing exhibit conventional spreading with strong adhesion. By capturing the fundamental interplay among electrostatic driving forces, contact-line pinning, and interfacial mobility, our results provide a new paradigm for controlled droplet transport and ejection in electrowetting systems mediated by dense micro-posts and lubricant-induced interfaces.

1. **Introduction**

Electrowetting-on-dielectric (EWOD) provides a powerful means of electrically controlling droplet wettability and motion, enabling applications in digital microfluidics, adaptive optics, and lab-on-a-chip platforms[1-2]. Classical descriptions of electrowetting rely on the Young–Lippmann relation, which assumes a rigid, smooth dielectric substrate and uniform electric fields[3]. While this framework successfully captures contact angle modulation on conventional

**Corresponding Author**
Arindam Das*, Associate Professor, School of Mechanical Sciences, Indian Institute of Technology (IIT) Goa, Email: arindam@iitgoa.ac.in

substrates, it becomes inadequate when extended to soft, microstructured, or liquid-infused interfaces[4].

Soft polymeric substrates, such as polydimethylsiloxane (PDMS), deform under capillary and electrical stresses, forming elastocapillary wetting ridges at the three-phase contact line[5-6]. These deformations introduce additional dissipation pathways and enhance contact-line pinning, significantly altering droplet dynamics under applied voltage. Simultaneously, surface microtextures introduce geometric confinement, field localization, and partial wetting states that further complicate electrowetting behavior. The combined influence of elasticity and texture can therefore lead to nonclassical spreading dynamics and electrowetting saturation.

Liquid-infused surfaces (LIS) represent a fundamentally different class of interfaces, where a lubricating liquid replaces solid–liquid contact at the interface[7,8]. Such surfaces exhibit extremely low contact-angle hysteresis and suppressed pinning, enabling smooth droplet motion and enhanced reversibility[9]. Recent studies on electrowetting on liquid-infused films have demonstrated reduced oscillations and improved actuation performance[10]; however, the dynamic instability modes arising from the interplay of lubrication, elasticity, and electrowetting remain poorly understood.

Here, we report a strikingly counterintuitive electrowetting response: droplets on certain microstructured and LIS substrates undergo rapid tangential motion and, in some cases, full ejection under DC voltages. Classical electrowetting theory predicts that enhanced wetting, facilitated by reduced contact angle, should favour spreading rather than droplet detachment. Our findings demonstrate that when pinning resistance is minimized, either through dense micro texturing supporting a Cassie state or via a continuous lubricant layer, minor asymmetries in electric stress distribution can produce lateral electrocapillary forces sufficient to overcome surface tension and viscous damping, leading to impulsive droplet motion. These findings provide a framework for understanding droplet ejection in low-pinning electrowetting environments and highlight the potential of LIS and engineered topologies for droplet actuation, beyond conventional spreading-controlled paradigms.

## 2. Experimental Methodology

### 2.1. Materials and Methods

For the dielectric, PDMS (Sylgard 184) was prepared using base-to-crosslinker ratios of 10:1 and 30:1 to obtain elastomers with approximate Young's moduli of ~1.5 MPa and ~60 kPa, respectively[11]. The mixtures were degassed and cast onto silicon masters to fabricate smooth and microtextured substrates via soft lithography, as shown in Figure 1 below. Square-post microtextures with post spacings of 5, 10, 20, 30, and 50 μm were used.

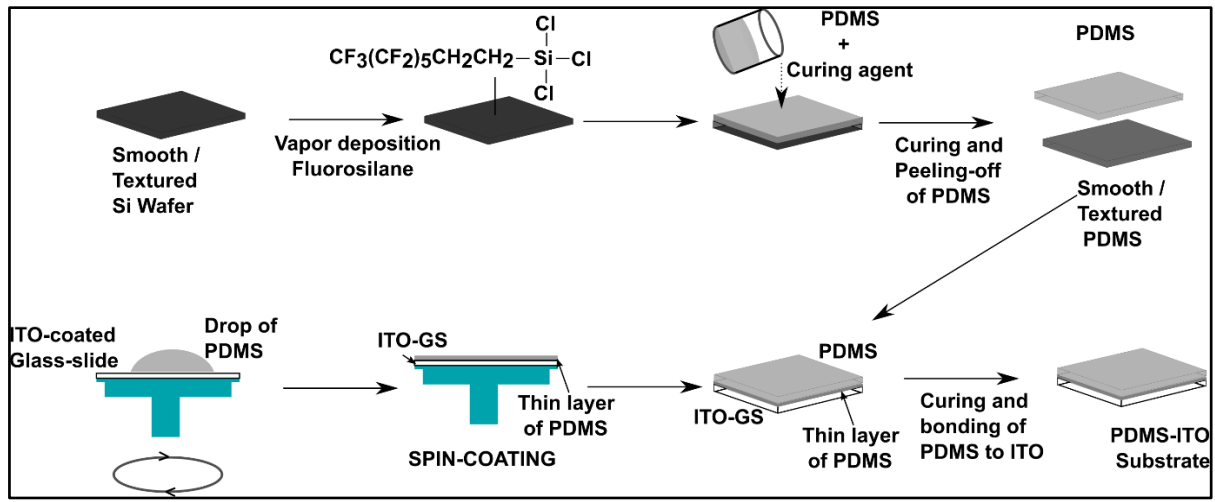

Figure 1 Schematic outlining of the fabrication of PDMS substrates and their subsequent bonding to the underlying ITO-coated glass, forming the dielectric–electrode assembly for electrowetting experiments.

All PDMS surfaces were rendered hydrophobic and chemically uniform by vapor-phase deposition of a fluorosilane monolayer (1H,1H,2H,2H-Perfluorooctyl-trichlorosilane - FS, Sigma-Aldrich). The treated PDMS layers were bonded to indium tin oxide (ITO)-coated glass substrates by spin-coating a thin PDMS adhesive layer, forming a dielectric–electrode assembly suitable for electrowetting experiments.

Textured PDMS substrates were infused with SO-5 cSt silicone oil to create lubricant-infused surfaces. The oil was selected based on immiscibility with water, strong wetting on fluorosilanized PDMS, and favorable spreading coefficient[12]. Lubricant infusion was achieved via dip-coating and drop-casting followed by controlled drainage, ensuring oil retention within the microtexture. Stability criteria based on critical contact angle and spreading coefficient confirmed the formation of a van der Waals–dominated LIS.

### 3. Electrowetting setup

Electrowetting experiments were conducted using a DC power supply with voltages from 1500V up to 5000V applied between the droplet (top electrode) and the grounded ITO substrate, as shown in Figure 2. Droplet dynamics were recorded using suitable high-speed imaging (1000/3000 fps), enabling measurement of contact angle evolution, wetted radius, and transient instability events. The captured images were used to evaluate parameters such as the apparent contact angle, contact radius, and spreading behaviour under an applied DC voltage.

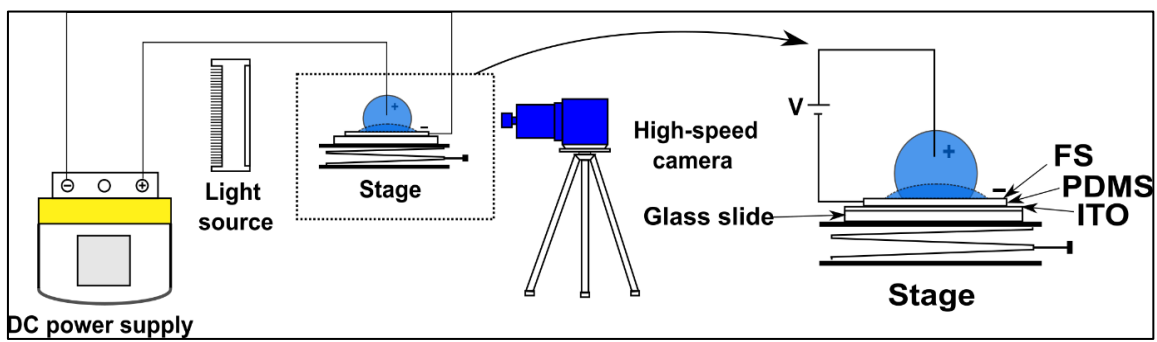

Figure 2 Schematic representation of the electrowetting experimental setup, showing the DC power supply, electrode configuration, illumination source, PDMS-coated ITO substrate mounted on the stage, and the high-speed imaging system used to capture droplet deformation under applied DC voltage.

## 4. Results and Discussion

### 4.1. Surface Wettability

Fluorosilanization increased the static water contact angle of smooth PDMS from ~108° to ~116° as shown in Figure 3. Microtexturing further amplified apparent contact angles, with densely packed textures (5–10 μm spacing) exhibiting Cassie-like behavior [13] and reduced contact-angle hysteresis (Figure 4 and Figure 5). Increasing post spacing led to a transition toward Wenzel-like wetting, accompanied by higher hysteresis. In addition, a slightly higher hysteresis was seen in softer samples.

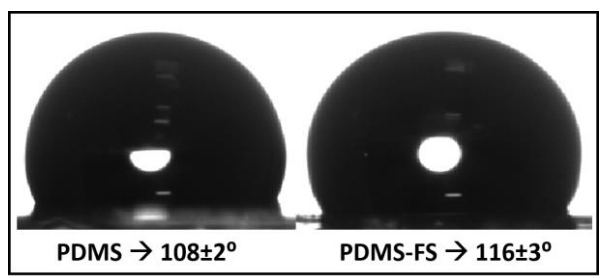

Figure 3 Equilibrium contact angles of bare PDMS and fluorosilanized PDMS (PDMS-FS) surfaces, illustrating the effect of surface chemical functionalization on wettability.

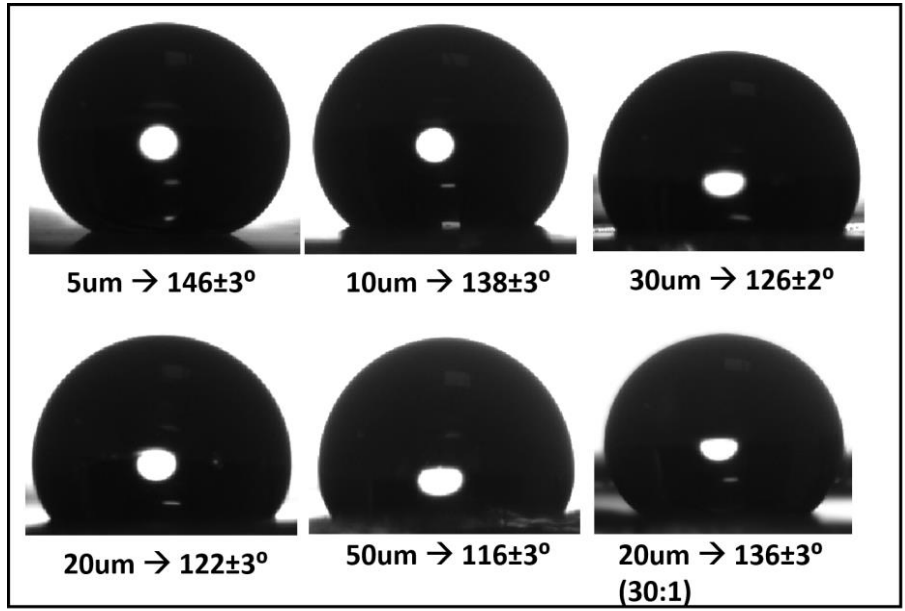

Figure 4 Equilibrium contact angles of fluorosilanized, microtextured PDMS surfaces with different post spacings, illustrating the influence of surface microtexture geometry on wettability.

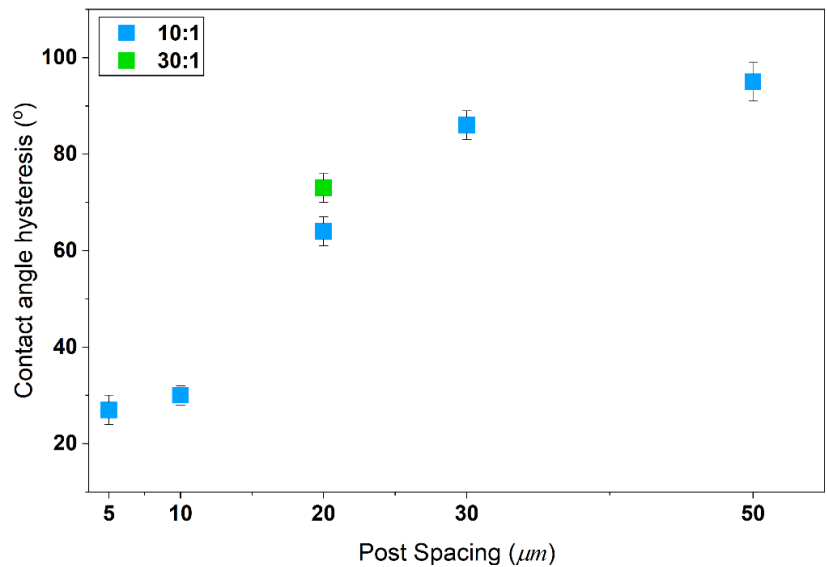

Figure 5 Contact angle hysteresis of PDMS samples with varying post spacings and surface functionalization, highlighting the combined effects of microtexture geometry and surface chemistry on contact-line pinning.

Lubricant-infused surfaces exhibited near-zero roll-off angles and suppressed hysteresis, confirming the presence of a continuous oil layer mediating droplet–substrate interactions. Spreading coefficient analysis showed that SO-5cSt silicone oil cloaks water droplets, supporting stable lubrication during electrowetting.

In the water environment, the contact angle of SO-5ct was measured at 22±3°. Table 1 and Table 2 implies that the silicone oil spreads readily in fluorosilanized and bare PDMS and that water exhibit moderate contact angle as well as low hysteresis in lubricant-coated PDMS.

**Table 1** Equilibrium contact angle of the lubricant(°) on the bare and fluorosilanized surfaces

| Samples → | PDMS | PDMS-FS |
|---|---|---|
| SO-5cst | 1±0.5 | 1±0.5 |

**Table 2** Wettability of bare and functionalised coated PDMS in SO-5cSt oil (°).

| Samples | PDMS | PDMS-FS |
|---|---|---|
| Eq. CA | 102 ± 3 | 89 ± 3 |
| ACA | 105 ± 2 | 89 ± 2 |
| RCA | 85 ± 3 | 78 ± 2 |
| CAH | 20 ± 5 | 11 ± 4 |

As mentioned earlier, dense post arrays trap air and yield high apparent hydrophobicity, while sparse posts allow the droplet to sag into the texture and wet more of the surface, reducing the macroscopic contact angle. This data is supported by quantification of air-layer stability within the texture, as explained below. The texture geometry used for the study is shown in the block diagram below.

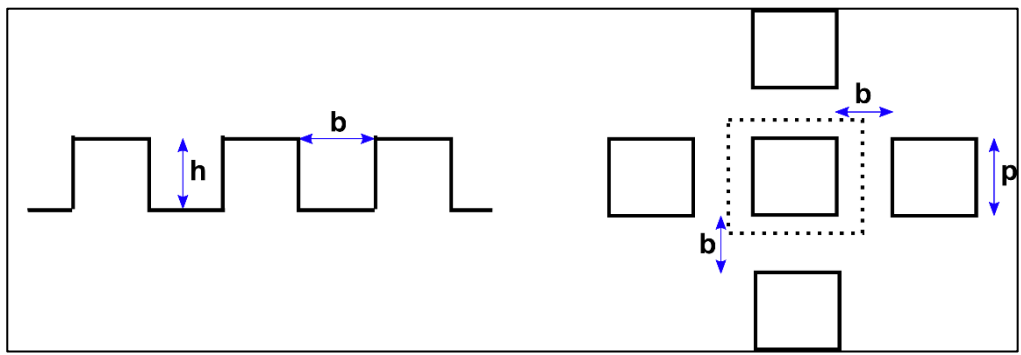

Figure 6 Schematic illustration of the microtexture geometry used in the study, showing the square post size (p), post spacing (b), and post height (h).

The total interfacial energy between air (a), water (w) and the silanized surface (FS) for Wenzel state is given by $(p^2 + 2bp + 4hp)\gamma_{FS-w}$

The total interfacial energy between air (a), water (w) and the silanized surface (FS) for Cassie state is given by $[(p + b)^2 - p^2]\gamma_{a-w} + (4hp + 2bp)\gamma_{FS-a}$

The water-fluorosilanized surface interfacial energy, $\gamma_{FS-w}$ was calculated from Young's equation.

Now solving for the condition that if the total interfacial energy between air (a), water (w) and the silanized surface (FS) for the Cassie state is less than that for the Wenzel state, then the Cassie state is said to be more favourable. The condition obtained was that, for a favourable Cassie state, the texture post-spacing should be less than 12.5 µm, which applies only to the dense post-spacing samples (5 and 10 µm).

### 4.2. LIS Stability

The characterization of the lubricant-infused surfaces (LIS) commenced with an assessment of their fundamental interfacial behaviours, particularly the tendency of the infused oil to cloak water droplets and the resulting implications for film stability. To probe this phenomenon, spreading coefficients[12] ($S_{ow(a)} = \gamma_{wa} - \gamma_{wo} - \gamma_{oa}$) were calculated using the measured interfacial tensions of SO-5cSt silicone oil (Table 3). These calculations revealed that SO-5cSt exhibits a positive spreading coefficient ($S_{ow(a)} > 0$), signifying that the lubricant spontaneously spreads over the water–air interface. This behaviour indicates the formation of a thin oil layer that cloaks or envelops the droplet, as shown in Figure 7, confirming that SO-5cSt is capable of cloaking the water droplet and contributing to the characteristic wetting response observed on LIS.

Table 3 Summary of the water–oil interfacial tension measurements and the associated spreading coefficients derived from them.

| Lubricant | $\gamma_{wa}$ (mN/m) | $\gamma_{wo}$ (mN/m) | $\gamma_{oa}$ (mN/m) | $S_{ow(a)}$ (mN/m) |
|---|---|---|---|---|
| Silicone oil (5cSt) | 71.83 | 47 | 19.7 | 5.3 |

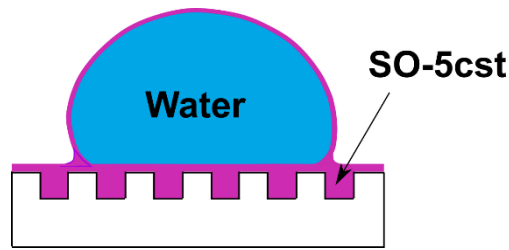

Figure 7 Schematic representation of a water droplet interacting with a lubricant-infused textured substrate, illustrating droplet cloaking by the surrounding lubricant layer (shown in purple) and the resulting liquid-mediated interfacial configuration.

Roll-off angle measurements were conducted to assess whether a continuous lubricating film of silicone oil was present above the micro-posts on both the smooth PDMS and textured

PDMS LIS surfaces. The SO-5 cSt–infused samples consistently exhibited markedly lower roll-off angles compared to their textured counterparts. This low roll-off angle value (< 5 °) indicates that the droplet primarily interacted with a stable oil layer rather than with the underlying solid features. The presence of this thin intervening film minimizes pinning at the contact line, thereby reducing resistance to droplet motion. Such behaviour is characteristic of a van der Waals–dominated LIS[14], where strong oil–substrate affinity supports the formation and retention of a continuous lubricating layer.

### 4.3. Electrowetting

#### 4.3.1. Deviation from Young–Lippmann Behavior

Measured contact angle variations under applied voltage deviated significantly from Young–Lippmann predictions on soft and textured substrates. These deviations increased with decreasing elastic modulus and increasing texture density, reflecting the influence of elastocapillary deformation, contact-line pinning, and electric-field localization near microfeatures.

On soft PDMS substrates, Maxwell stresses induce electromechanical coupling, resulting in substrate deformation and the formation of a wetting ridge at the contact line, which leads to pinning and hysteresis that hinder droplet spreading[15]. The effect is amplified in softer (30:1) samples due to reduced elastic modulus, which enhances viscoelastic dissipation during contact line motion. Moreover, textured substrates introduce additional pinning sites and surface heterogeneity, further intensifying deviations and accelerating the saturation of contact angles. These cause deviations as shown in Figure 8 establishing that, while the Lippmann–Young equation captures the idealised electrowetting response, real systems on soft and structured dielectrics are governed by complex interfacial mechanics. The findings underscore the need to incorporate substrate elasticity, geometry, and viscoelastic energy dissipation into extended models to predict electrowetting behaviour on functional soft surfaces.

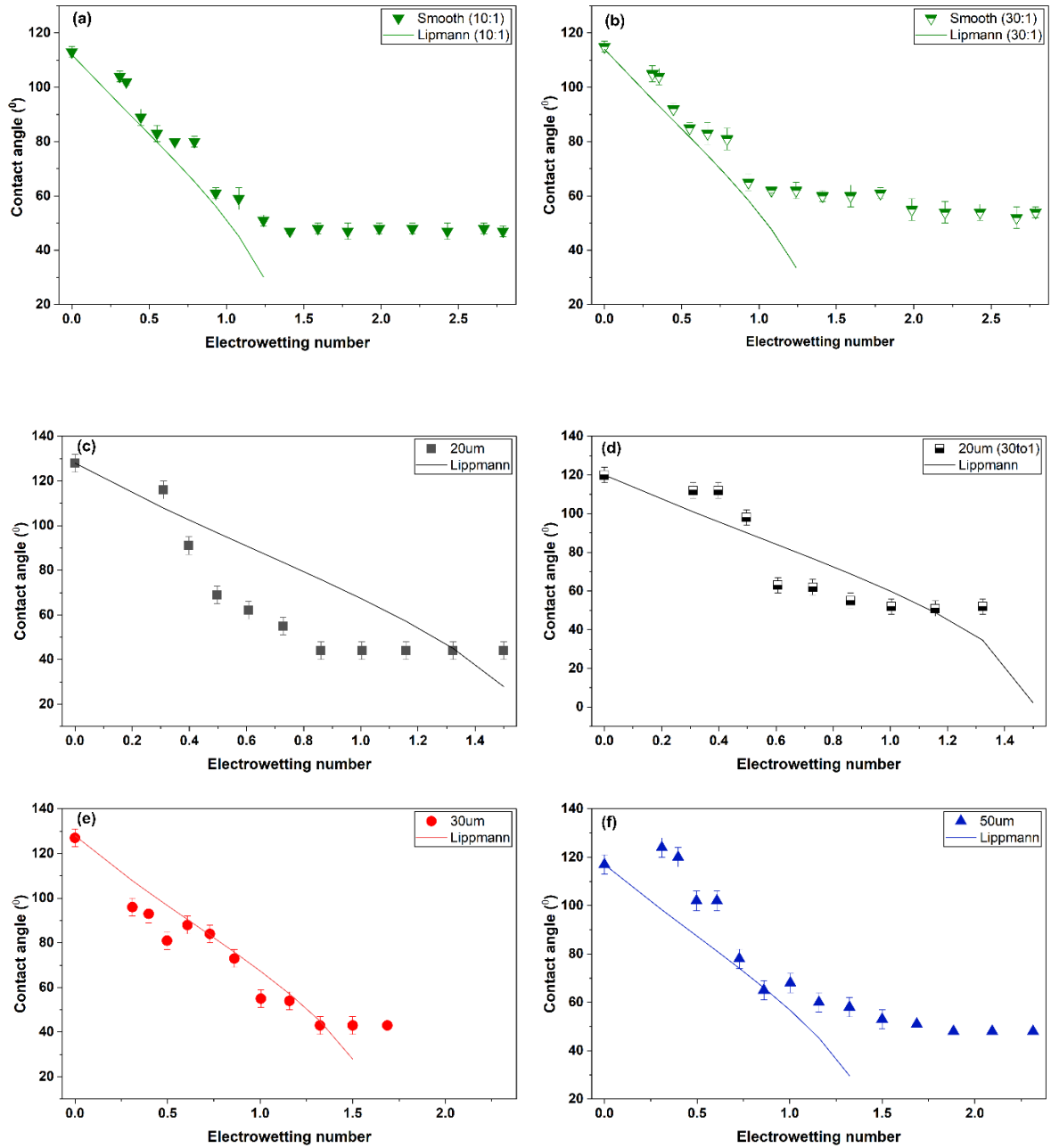

Figure 8 Equilibrium contact angle as a function of electrowetting number for PDMS substrates with varying elasticity and surface textures, compared with the predicted Lippmann–Young relation for (a) smooth PDMS (10:1), (b) smooth PDMS (30:1), (c) 20 μm textured PDMS (10:1), (d) 20 μm textured PDMS (30:1), (e) 30 μm textured PDMS, and (f) 50 μm textured PDMS surfaces.

The samples assessed in this study exhibited two qualitatively different response modes.: Electrowetting-Induced Droplet Ejection and Pinning-limited spreading.

### 4.3.2. Electrowetting-Induced Droplet Ejection

Electrowetting experiments performed on fluorosilanized and lubricant-infused PDMS substrates revealed that ejection of droplet occurred on a specific subset of surfaces: the micro-textured PDMS samples with 5 μm and 10 μm post spacings in their fluorosilanized state, and all silicone-oil–infused textured substrates, regardless of spacing. The mechanisms underlying these contrasting outcomes are described in detail below.

The time-resolved snapshots of the spreading and ejection in the dense post-microstructured (5 μm and 10 μm) surfaces are shown in Figure 9.

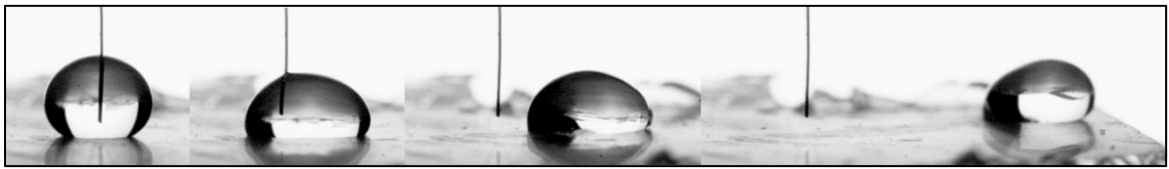

Figure 9 Sequential snapshots showing the temporal evolution of electrowetting-induced ejection of a droplet on densely confined 5 μm microstructured PDMS. The images illustrate the deformation of the droplet interface, the expansion of the wetted footprint, and the subsequent ejection of the droplet. It shows how strongly the wetting dynamics change under an applied electric field.

In the present system, macroscopic electrowetting is strongly suppressed by the use of a thick dielectric layer (1.6 mm), which significantly reduces the electrowetting number and inhibits uniform contact-angle modulation or classical Cassie–Wenzel transitions. Consequently, the droplet response is governed not by global wetting transitions, but by local interfacial processes controlled by surface texture and capillarity.

When a voltage is applied in an electrowetting configuration involving a water droplet on PDMS, electric stresses develop predominantly near the droplet-solid interface and the three-phase contact line due to leaky dielectric behavior and interfacial charge relaxation[16]. These stresses reduce the effective solid-liquid interfacial energy and promote electrohydrodynamic flow along the interface, consistent with established electrowetting and electrohydrodynamic frameworks[16]. On low-pitch microtextures, droplets adopt a Cassie-like state, reducing solid–liquid contact and contact-line hysteresis. When a voltage is applied to these surfaces, electric stresses concentrate at the three-phase contact line, and microstructural heterogeneities cause the droplet to move downward, creating a contact angle imbalance due to small surface asymmetries. When voltage is applied to the imbalanced droplet lateral electrocapillary pressure gradients are induced that can propel the droplet out of the system.

The characteristic time scale of this event (≈0.1–0.15 s) is consistent with an instability-driven process rather than quasi-static spreading. At higher voltages, partial vertical lifting is occasionally observed, accompanied by necking and breakup of the droplet from the surface, resembling the Leidenfrost effect[17] seen during droplet impact on hydrophobic substrates.

When the microtextured PDMS surfaces were impregnated with SO-5cSt silicone oil, droplet ejection was observed across all geometries, regardless of post spacing or substrate stiffness. The snapshots of the ejection are shown in Figure 10 below.

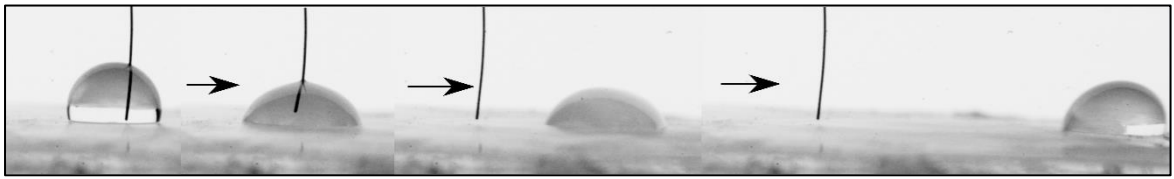

Figure 10 Sequential snapshots showing the temporal evolution of electrowetting-induced ejection on lubricant-impregnated PDMS microtextured surfaces. The images illustrate the smooth interface deformation and continuous expansion followed by ejection of the wetted footprint, highlighting the role of the infused lubricant in suppressing pinning and enabling friction-reduced wetting under an applied electric field.

Here, the continuous lubricant layer effectively eliminates contact-line pinning. Even small asymmetries in the electric field or substrate geometry generate lateral forces. Upon voltage application, electric stresses localise at the curved water-lubricant interface, leading to a downward movement and an imbalance, particularly at the advancing front of the droplet, where it is asymmetrically deformed. This asymmetric droplet, on the addition of more charge, induces electrohydrodynamic flow within the droplet, causing the advancing edge to elongate and move forward. As the leading-edge advances, the trailing edge remains temporarily stationary due to viscous resistance within the lubricant ridge or cloaking oil layer. This produces a transient asymmetry in interfacial curvature and stress distribution. Electrostatic and interfacial energy therefore accumulates until the resisting viscous forces at the receding end are overcome, at which point the trailing edge is abruptly pulled forward. Repetition of this deformation-release cycle gives rise to an inchworm–like crawling motion, characterized by discrete, stepwise advancement of the droplet.

The characteristic timescale of this crawling motion is approximately 0.2 s, which is slightly longer than the timescale observed on low-pitch dry textured PDMS surfaces (~0.1–0.15 s). This increase in timescale reflects the additional viscous dissipation and delayed stress relaxation associated with the lubricant ridge, as opposed to the primarily pinning-controlled dynamics on dry microtextures. At higher applied voltages, the magnitude of interfacial deformation and the accumulated electrostatic energy increase further, leading to a transition from crawling to sudden lateral droplet ejection, which occurs over a similarly short, instability-driven timescale.

In contrast, droplets in the Wenzel state or on high-pitch textured surfaces with strong pinning dissipate force imbalances through droplet deformation and re-anchoring. As a result, ejection is prevented.

### 4.3.3. Pinning-limited spreading.

On other substrates with higher post-spacing (20, 30, and 50 μm) and on a soft substrate, ejection did not occur due to strong pinning; instead, the droplet spread in a controlled fashion (or remained stuck), and the final wetted radius was smaller.

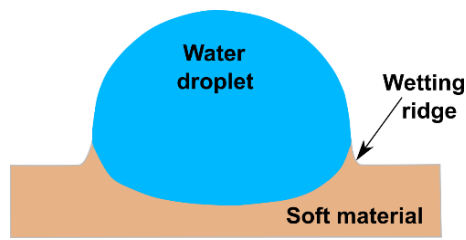

Figure 11 Schematic illustration of a wetting ridge formed at the three-phase contact line on a soft substrate, highlighting elastocapillary deformation induced by surface tension forces.

The 30:1 PDMS is substantially softer (lower E) than the 10:1 PDMS; softer substrates allow a more pronounced wetting ridge, as shown in Figure 11, to form at the contact line under capillary and electrical stresses, making it the highest-pinning sample[11,18]. This ridge increases the effective contact-line adhesion and hysteresis (pinning), so the net depinning force required is larger. Pinning manifests as contact-angle hysteresis and intermittent stick–slip of the contact line during voltage application. The observed correlation between substrate compliance and pinning (softer substrates often giving stronger pinning through ridge formation) matches recent observations in soft electrowetting[11]. For all textured samples with a 10:1 curing ratio, they behaved like rigid samples, with little effect from the material's softness.

The substrate also has a moderate level of contact-angle hysteresis, which dampens contact-line motion and stabilizes the wetted footprint of the droplet. As a result, electrowetting-induced deformation proceeds in a controlled and continuous manner. Without the confinement required for energy accumulation and sudden release, the droplet remains firmly in the spreading regime and never undergoes ejection driven by instability, as was observed in high-speed imaging shown in Figure 12 below.

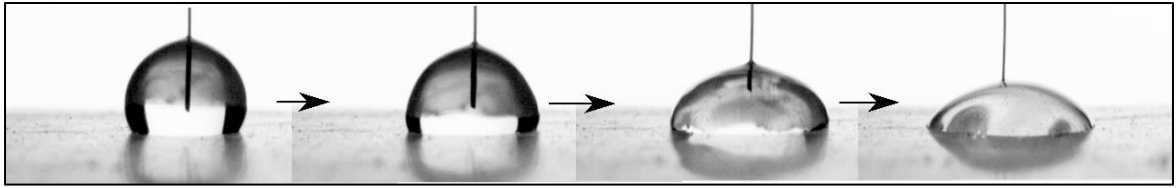

Figure 12 Sequential snapshots showing the temporal evolution of electrowetting-induced spreading on pinning-dominated PDMS surfaces, including smooth and large-pitch microtextured samples. The images highlight the arrested movement of the contact line.

Further, for each sample, the temporal evolution of the wetted base radius, $R_c(t)$, of sessile deionised-water droplets was measured for droplets subjected to a DC voltage applied through a top electrode (positive) and a grounded electrode beneath the PDMS dielectric. From the raw data we computed (i) the dimensional radius evolution $R_c(t)$, and (ii) dimensional curves $R'_c(t^*)$ using scaling by non-dimensionalising it with the initial value $R'_c = \frac{R_c(t)}{R_c^0}$, where $R_c^0$ is the initial equilibrium contact radius.

### 4.3.3.1. Radius evolution and nondimensional behaviour

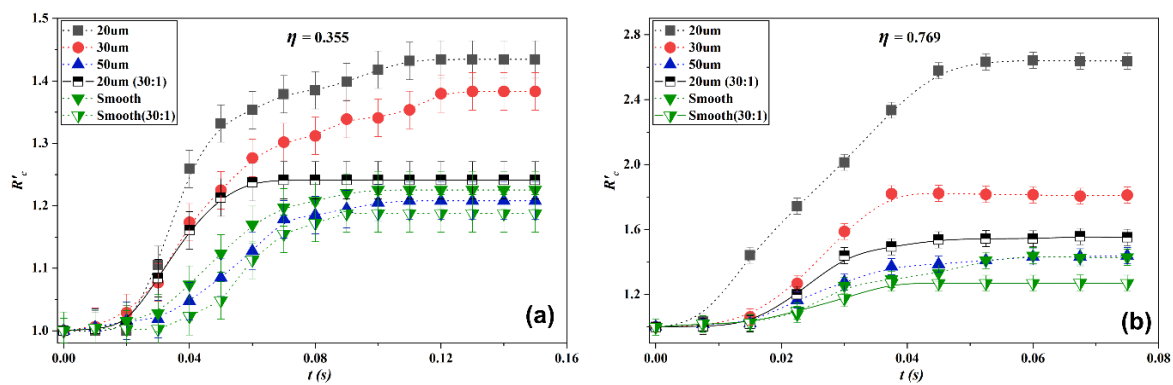

Figure 13 Temporal evolution of contact radius for constant electrowetting number (for 1700V and 2500V) for PDMS substrates of varying elasticity and surface textures.

The nondimensional radius evolution $R'_c$, as a function of time, $t$, for samples that exhibited pinning, is plotted in Figure 13. The 20 μm sample (10:1) exhibited the most advanced spreading for a given $t$, i.e., the least effective pinning, producing the largest $R'_c$ at fixed nondimensional time. At early stages (immediately after voltage application), the wetted/contact radius shows a faster transient (inertial/electric-field-driven) increase on low-pinning substrates, often followed by a viscous reconfiguration. On high-pinning samples (smooth 30:1), the initial growth of $R'_C$ is arrested (constant-radius regimes) until depinning events occur, which showed the slowest overall dynamics and the largest residence time in the pinned (constant-radius) plateaus. These transient modes mirror the standard stages of droplet spreading (early slow viscous stage → faster non-hysteretic regime) modified by the electrowetting forcing and the presence of microscale texture.

To quantify the influence of substrate compliance on voltage-induced droplet spreading, an amplification factor was defined as the ratio of the normalized final contact radius on a rigid substrate to that on a soft substrate, $A = \frac{(R'_C)_{rigid}}{(R'_C)_{soft}}$, evaluated separately for both textured and smooth surface geometries. For an electrowetting number of 0.355, the amplification factor was 1.17 for textured surfaces and 1.03 for smooth surfaces, indicating a pronounced reduction in contact-line expansion on soft textured substrates compared to their rigid counterparts. The other values are displayed in Table 4 and Table 5 below. This disparity suggests that surface texture amplifies the role of substrate compliance under electrowetting conditions, likely due to enhanced electric-field localization and increased contact-line pinning, which together restrict radial spreading on deformable substrates. At a higher electrowetting number of 0.769, the amplification factors increased to 1.733 for textured surfaces and 1.094 for smooth surfaces, reflecting a stronger voltage dependence of the compliance effect. The substantial rise in amplification for textured surfaces at higher electrowetting numbers suggests that increasing electrostatic forcing intensifies mechanical–electrical coupling at the contact line, thereby leading to greater suppression of spreading on soft substrates. In contrast, the consistently low amplification observed for smooth surfaces across both electrowetting numbers suggests that, in the absence of texture-induced field gradients and pinning sites, substrate softness plays a relatively minor role in governing the final contact radius. Overall, these results (Table 4 and Table 5) demonstrate that substrate compliance significantly modulates electrowetting-induced spreading in a texture-dependent manner, with its influence becoming increasingly prominent at higher applied voltages.

Table 4 Amplification factor for textured surfaces as a function of applied voltage and electrowetting number.

| Electrowetting number | Voltage (V) | Amplification factor (A) |
|---|---|---|
| 0.2 | 1250 | 1.07 |
| 0.355 | 1700 | 1.17 |
| 0.5 | 2000 | 1.3 |
| 0.65 | 2300 | 1.53 |
| 0.769 | 2500 | 1.73 |

Table 5 Amplification factor for smooth surfaces as a function of applied voltage and electrowetting number

| Electrowetting number | Voltage (V) | Amplification factor (A) |
|---|---|---|
| 0.2 | 1250 | 1.01 |
| 0.355 | 1700 | 1.03 |
| 0.5 | 2000 | 1.053 |
| 0.65 | 2300 | 1.072 |
| 0.769 | 2500 | 1.09 |

These trends indicate that the textured geometry strongly modifies pinning and, thus, the functional form, using all the dependent parameters derived from the Buckingham pi theorem, is reported below.

$$R_c = f\left(\frac{tE}{\mu}, \frac{\gamma}{Es}, \eta\right)$$

Where $E$ is the Young's modulus of elasticity of the PDMS substrate, s = characteristic post spacing (μm), t = time, μ = dynamic viscosity of the liquid, γ = liquid surface tension.

This representation clarifies which combinations of material, geometric, and electrical parameters govern the evolution of the radius through dimensionless numbers.

γ/($E$s) (the elastocapillary parameter) controls whether the substrate deforms significantly at the contact line: when γ/($E$s) is large (soft substrate, large γ or small s) substrate deformation and ridge formation are important and tend to increase pinning; when it is small (stiff substrate or large spacing) the substrate behaves effectively rigid on the texture length, and pinning from substrate deformation is reduced. Here, manifesting it in a different way without counting the post spacing 's', calculating $\gamma/E$ can be called as a characteristic length scale. For the softer PDMS (30:1), the length scale is 1.2 μm, whereas it is 48nm for the 10:1 PDMS. So, it is clear that the softness effect will be on the 30:1 PDMS.

The nondimensional time $tE/\mu$ groups the roles of substrate elasticity and liquid viscosity in the dynamics: increasing E or t, or decreasing μ, increases $tE/\mu$ and typically advances the spreading. $tE/\mu$ compares the observation time to a characteristic viscous–elastic coupling timescale $\mu/E$.

The electrowetting number is the dimensionless measure of applied voltage for dielectric-coated substrates, as explained in the earlier sections.

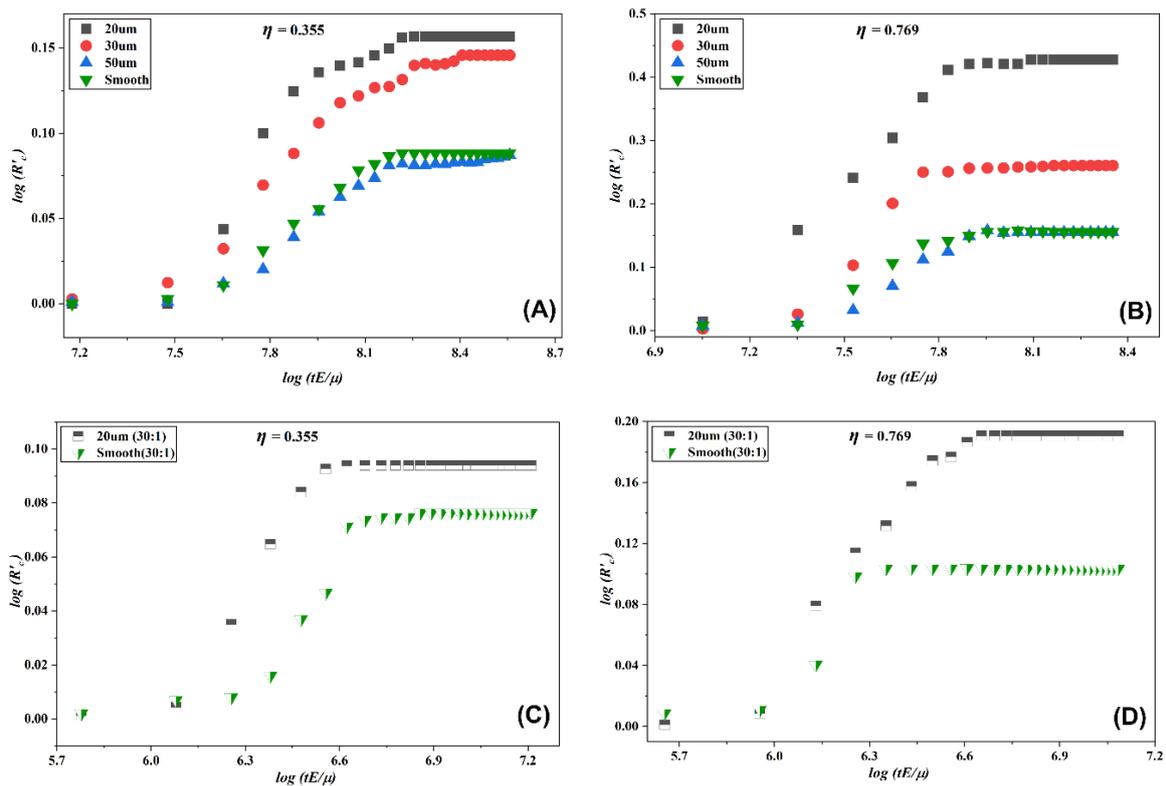

Figure 14 Temporal evolution of the wetted contact radius for PDMS substrates of different elasticities under fixed electrowetting numbers corresponding to applied voltages of 1700 V and 2500 V, illustrating the influence of substrate stiffness on electrowetting-driven spreading dynamics.

Figure 14 shows the log plot of $R_c'$ plotted against nondimensional time at fixed electrowetting numbers for PDMS substrates of different elasticities. For rigid samples, motion begins once

$tE/\mu \approx 7.2$, marking the point where electrical forcing exceeds the initial frictional threshold. When $tE/\mu > 7.2$, the droplet transitions into a rapid spreading regime dominated by viscous–elastic interactions. For softer substrates, this threshold decreases to $tE/\mu \approx 5.7$, consistent with their lower elastic resistance. The y-axis (log plot of wetted radius), which represents the degree of spreading, indicates that spreading is greater on rigid surfaces because they offer less resistance from elastic interactions. These results collectively demonstrate that substrate elasticity, microscale geometry, and viscoelastic dissipation jointly dictate electrowetting spreading dynamics. Soft substrates introduce additional energy loss pathways and deformation-driven pinning, whereas microtextures modulate the force balance through geometric confinement. The combination of elasticity-dependent pinning, texture-induced barriers, and voltage-driven forcing yields a rich spectrum of dynamic behaviors. Incorporating these parameters into extended theoretical models is therefore essential for accurately predicting electrowetting responses on engineered soft surfaces.

## 5. Conclusion

We demonstrate that electrowetting on soft, microstructured, and lubricant-infused PDMS interfaces produces dynamic responses that depart fundamentally from classical Young–Lippmann behavior. By systematically varying substrate elasticity, texture pitch, and lubricant presence, two distinct regimes emerge: pinning-limited spreading and instability-driven droplet ejection. Dense microtextures that support Cassie states induce an asymmetry in electrocapillary forces, culminating in rapid lateral ejection. In contrast, larger-pitch and soft substrates enhance elastocapillary pinning, suppressing instabilities and favoring controlled spreading. Lubricant-infused surfaces eliminate solid–liquid pinning altogether, promoting asymmetric electrohydrodynamic deformation and inchworm-like translational motion prior to detachment.

A unified dimensionless framework incorporating elastocapillary effects (γ/Es), viscosity–elastic coupling (tE/μ), geometric confinement (R/s), and electrowetting number captures the hierarchy of observed behaviors in the sparse post microstructures. These findings establish that coupled electrostatics, elasticity, texture, and interfacial mobility, not voltage alone, govern droplet transport on functional soft interfaces. The results provide insight and design guidelines for electrowetting-enabled droplet propulsion, programmable transport, and adaptive microfluidic systems on deformable surfaces.

# REFERENCES

1   X. Liu, D. Ma, H. Ye, Y. Hou, X. Bai, Y. Xing, X. Cheng, B. Lin and Y. Lu, Electrowetting-based digital microfluidics: Toward a full-functional miniaturized platform for biochemical and biological applications, *TrAC Trends in Analytical Chemistry*, 2023, **166**, 117153.

2   Biofunctionalization of electrowetting-on-dielectric digital microfluidic chips for miniaturized cell-based applications - Lab on a Chip (RSC Publishing), https://pubs.rsc.org/en/content/articlelanding/2011/lc/c1lc20340a, (accessed 13 November 2025).

3   F. Mugele and J.-C. Baret, Electrowetting: from basics to applications, *Journal of Physics: Condensed Matter*, 2005, **17**, R705–R774.

4   I. E. Markodimitrakis, D. G. Sema, N. T. Chamakos, P. Papadopoulos and A. G. Papathanasiou, Impact of substrate elasticity on contact angle saturation in electrowetting, *Soft Matter*, 2021, **17**, 4335–4341.

5   Dynamics of droplet spreading on a flexible substrate under electrowetting-on-dielectric | Physics of Fluids | AIP Publishing, https://pubs.aip.org/aip/pof/article-abstract/37/10/102012/3367450/Dynamics-of-droplet-spreading-on-a-flexible?redirectedFrom=fulltext, (accessed 12 November 2025).

6   Wetting on silicone surfaces - Soft Matter (RSC Publishing) DOI:10.1039/D4SM00346B, https://pubs.rsc.org/en/content/articlehtml/2024/sm/d4sm00346b, (accessed 12 November 2025).

7   Contact-angle hysteresis provides resistance to drainage of liquid-infused surfaces in turbulent flows | Phys. Rev. Fluids, https://journals.aps.org/prfluids/abstract/10.1103/PhysRevFluids.9.054002, (accessed 13 November 2025).

8   Evaporation of Sessile Droplets on Slippery Liquid-Infused Porous Surfaces (SLIPS) | Langmuir, https://pubs.acs.org/doi/10.1021/acs.langmuir.5b03240, (accessed 12 November 2025).

9   Electrowetting on liquid-infused film (EWOLF): Complete reversibility and controlled droplet oscillation suppression for fast optical imaging | Scientific Reports, https://www.nature.com/articles/srep06846, (accessed 13 November 2025).

10  Lubrication effects on droplet manipulation by electrowetting-on-dielectric (EWOD) | Journal of Applied Physics | AIP Publishing, https://pubs.aip.org/aip/jap/article-abstract/132/20/204701/2837857/Lubrication-effects-on-droplet-manipulation-by?redirectedFrom=fulltext, (accessed 13 November 2025).

11  Electrically modulated dynamic spreading of drops on soft surfaces | Applied Physics Letters | AIP Publishing, https://pubs.aip.org/aip/apl/article-abstract/107/3/034101/133466/Electrically-modulated-dynamic-spreading-of-drops?redirectedFrom=fulltext, (accessed 18 September 2025).

12  J. D. Smith, R. Dhiman, S. Anand, E. Reza-Garduno, R. E. Cohen, G. H. McKinley and K. K. Varanasi, Droplet mobility on lubricant-impregnated surfaces, *Soft Matter*, 2013, **9**, 1772–1780.

13  Fakir droplets | Nature Materials, https://www.nature.com/articles/nmat715, (accessed 15 November 2025).

14  High-Speed Imagery Analysis of Droplet Impact on van der Waals and Non-van der Waals Soft Oil-Infused Surfaces | Langmuir, https://pubs.acs.org/doi/full/10.1021/acs.langmuir.5c02765, (accessed 17 November 2025).

15  R. W. Style, Universal Deformation of Soft Substrates Near a Contact Line and the Direct Measurement of Solid Surface Stresses, *Phys. Rev. Lett.*, DOI:10.1103/PhysRevLett.110.066103.
16  D. A. Saville, ELECTROHYDRODYNAMICS: The Taylor-Melcher Leaky Dielectric Model, *Annual Review of Fluid Mechanics*, 1997, **29**, 27–64.
17  BOILING, https://www.sciencedirect.com/science/chapter/monograph/abs/pii/B9780123706102500155, (accessed 16 January 2026).
18  Mechanical properties of bulk Sylgard 184 and its extension with silicone oil | Scientific Reports, https://www.nature.com/articles/s41598-021-98694-2?utm_source=chatgpt.com, (accessed 18 September 2025).